\documentclass{iaus}
\usepackage{graphicx}

\title[Habitability of super-Earths] 
{Habitability of super-Earths:\\ Gliese 581c \& 581d}

\author[von Bloh et al.]{W. von Bloh$^1$, C. Bounama$^1$, M. Cuntz$^2$ \and S. Franck$^1$}

\affiliation{$^1$Potsdam Institute for Climate Impact Research,\\ P.O. Box 601203, Potsdam, Germany,
\\ email: {\tt bloh@pik-potsdam.de, bounama@pik-potsdam.de, franck@pik-potsdam.de} \\
[\affilskip]
$^2$University of Texas at Arlington,\\ P.O. Box 19059, Arlington, TX 76019, USA
\\email: {\tt cuntz@uta.edu}}

\pubyear{2008}
\volume{249}  
\pagerange{119--126}
\jname{Exoplanets: Detection, Formation \& Dynamics}
\editors{A.C. Editor, B.D. Editor \& C.E. Editor, eds.}
\begin{document}

\maketitle

\begin{abstract}
The unexpected diversity of exoplanets includes a growing number of super-Earth planets,
i.e., exoplanets with masses smaller than 10 Earth masses.   Unlike the larger exoplanets
previously found, these smaller planets are more likely to have similar chemical and
mineralogical composition to the Earth.  We present a thermal evolution model for
super-Earth planets to identify the sources and sinks of atmospheric carbon dioxide.
The photosynthesis-sustaining habitable zone (pHZ) is determined by the limits of
biological productivity on the planetary surface. We apply our model to calculate
the habitability of the two super-Earths in the Gliese 581 system. The super-Earth
Gl~581c is clearly outside the pHZ, while Gl~581d is at the outer edge of the pHZ.
Therefore, it could at least harbor some primitive forms of life.
\end{abstract}

\firstsection 
\section{Introduction}

Very recently, \cite[Udry et al. (2007)]{udry07}
announced the detection of two super-Earth planets in the Gliese 581 system; namely,
Gl~581c with a mass of
5.06~$M_\oplus$ and a semi-major axis of 0.073 AU, and Gl~581d with 8.3~$M_\oplus$ and
0.25 AU.  Both mass estimates are minimum masses uncorrected for $\sin i$. The luminosity
of Gl~581 can be estimated as $L=0.013 \pm 0.002 L_\odot$  with a stellar temperature of
$T_e= 3480$~K and a stellar age of at least 2 Gyr.
The main question is whether any of the two super-Earths around Gl~581 can harbor life,
i.e., that any of the planets lie within the habitable zone (HZ).  Typically, stellar HZs
are defined as regions around the central star, where the physical conditions are favorable
for liquid water to exist at the planet's surface for a period of time long enough for
biological evolution to occur (\cite[Kasting et al. 1993]{kasting93}).

In the following, we adopt a definition of the HZ previously used by
\cite[Franck et al. (2000)]{franck00a}.  Here habitability 
does not just depend on the parameters of the central star, but
also on the properties of the planet.  In particular, habitability is linked
to the photosynthetic activity of the planet, which in turn depends on the
planetary atmospheric CO$_2$ concentration, and is thus strongly influenced by
the planetary dynamics.  We call this definition the
photosynthesis-sustaining habitable zone (pHZ). In principle, this
leads to additional spatial and temporal limitations of habitability.

\section{Estimating the Habitability of a Super-Earth}

To assess the habitability of a super-Earth, i.e., a rocky planet smaller
than 10 Earth masses (\cite[Valencia et al. 2006]{valencia06}), an Earth-system model
is applied to calculate the evolution of the temperature and 
atmospheric CO$_2$ concentration.  
The numerical model couples the stellar luminosity, the silicate-rock weathering rate,
and the global energy balance to obtain estimates of the partial pressure of atmospheric
carbon dioxide $P_{\mathrm{CO}_2}$, the mean global surface temperature $T_\mathrm{surf}$,
and the biological productivity  $\Pi$ as a function of time $t$.  The main point is the
persistent balance between the CO$_2$ sink in the atmosphere-ocean system and the
metamorphic (plate-tectonic) sources.  This is expressed through the dimensionless
quantities
\begin{equation}
f_{\mathrm{wr}}(t) \cdot f_A(t) = f_{\mathrm{sr}}(t),
\label{gfr}
\end{equation}
where $f_{\mathrm{wr}}(t)$ is the 
weathering rate, $f_A(t)$ is 
the continental area, and $f_{\mathrm{sr}}(t)
$ is the spreading rate normalized by their present values of Earth. 
The connection between the stellar parameters and the planetary climate can be
formulated by using a radiation balance equation.
The evolution of the surface temperature is derived directly from the stellar
luminosity, the distance to the central star and the geophysical forcing ratio
($\mathrm{GFR} := f_{\mathrm{sr}} / f_A$). For the investigation of a super-Earth
under external forcing, we adopt a model planet with a prescribed continental area.
The fraction of continental area to the total planetary surface is varied between
$0.1$ and $0.9$. 

The thermal history and future of a super-Earth is determined by calculating the
GFR values. Spreading rates can be derived from the mantle temperature.
Assuming conservation of energy, the average mantle temperature $T_m$ is obtained as
\begin{equation} {4 \over 3} \pi \rho c (R_m^3-R_c^3) \frac{dT_m}{dt} = -4 \pi
R_m^2 q_m + {4 \over 3} \pi E(t) (R_m^3-R_c^3), \label{therm} \end{equation}
where $\rho$ is the density, $c$ is the specific heat at constant pressure,
$q_m$ is the heat flow from the mantle, $E$ is the energy production rate by
decay of radiogenic heat sources in the mantle per unit volume, and $R_m$ and
$R_c$ are the outer and inner radii of the mantle, respectively.
The photosynthesis-sustaining HZ (pHZ) is defined as the spatial domain encompassing
all distances $R$ from the central star where the biological productivity
is greater than zero, i.e.,
\begin{equation}
{\rm pHZ} := \{ R \mid \Pi (P_{\mathrm{CO}_2}(R,t), T_{\mathrm{surf}}(R,t))>0 \}.
\label{hz}
\end{equation}
In our model, biological productivity is considered to be solely a function of
the surface temperature and the CO$_2$ partial pressure in the atmosphere.
Our parameterization yields 
zero productivity for $T_{\mathrm{surf}} \leq 0^{\circ}$C or $T_{\mathrm{surf}}
\geq 100^{\circ}$C or $P_{\mathrm{CO}_2}\leq 10^{-5}$ bar
(\cite[Franck et al. 2000]{franck00a}).
To calculate the spreading rates for a planet with several Earth masses, 
the planetary parameters have been adjusted following
\cite[Valencia et al. (2006)]{valencia06} as
\begin{equation}
\frac{R_p}{R_{\oplus}}= \left(\frac{M}{M_{\oplus}}\right)^{0.27},
\end{equation}
where $R_p$ is the planetary radius and $M$ is the mass, with subscript $\oplus$
denoting Earth values.  See \cite[von Bloh et al. (2007)]{vonbloh07} for details
and specified parameter values.


\section{Results and Discussion}

\begin{figure}[h]
\centering
\resizebox{0.515 \hsize}{!}{\includegraphics{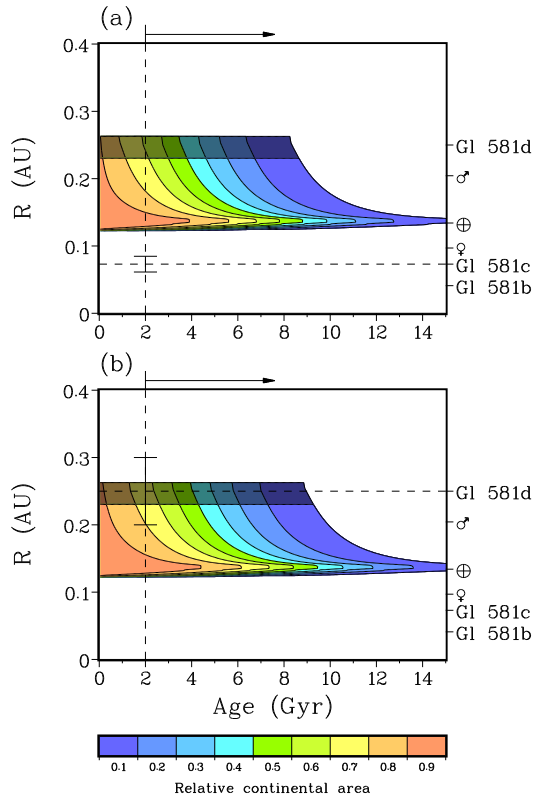}}
\caption{The pHz zone of Gl~581 for super-Earth Gl~581c (a) and Gl~581d (b) with a relative
continental area varied from $0.1$ to $0.9$ and a fixed stellar luminosity of 
$0.013 L_\odot$.  The results are given as a function of planetary age.
The light colors correspond to a maximum CO$_2$ pressure of 5 bar, whereas the
dark colors correspond to 10 bar.  For comparison, the positions of Venus, Earth
and Mars are shown scaled to the luminosity of Gl~581.}
\label{fig1}
\end{figure}

The habitable zone around Gl~581 for planets with five and eight Earth masses has been
calculated for
$L=0.013 L_\odot$.  The results for Gl~581c and Gl~581d  are shown in Fig. \ref{fig1}a,b.
The simulations
have been carried out for a maximum CO$_2$ pressure of 5 bar (light colors) and 10 bar
(dark colors) neglecting the cooling effect of CO$_2$ clouds.  
The super-Earth planet Gl~581c is found to be clearly outside the habitable zone.
On the other hand, one might expect that life had a chance to originate on Gl~581d
because it is near the outer edge of the pHZ, and for some combination of system
parameters even inside the pHZ (\cite[von Bloh et al. 2007]{vonbloh07}).
A planet of eight Earth masses has more volatiles than an Earth-size planet and can
build up a sufficiently dense atmosphere to prevent it from freezing out due to
tidal locking.  Planets inside the habitable zone around M stars may be tidally locked,
which however does not necessarily thwart habitability
(\cite[Tarter et al. 2007]{tar07}).  The modestly eccentric orbit of Gl~581d
($e=0.2\pm0.1$) further supports habitability, even if the
maximum CO$_2$ pressure is assumed to be as low as 5 bar.  The appearance of complex life,
however, is unlikely due to the rather adverse environmental conditions.  To get an
ultimate answer to
the profound question of life on Gl~581d, we have to await the TPF/Darwin missions.
They will allow for the first time the detection of biomarkers in the atmospheres of the
two super-Earths around Gl~581.

\end{document}